\documentclass[12pt]{article}
 \usepackage{epsfig}
 \def\be{\begin{equation}}
 \def\ee{\end{equation}}
 \def\bea{\begin{eqnarray}}
 \def\eea{\end{eqnarray}}
 \usepackage{graphicx}

 \catcode`\@=11
 \def\lsim{\mathrel{\mathpalette\@versim<}}
 \def\gsim{\mathrel{\mathpalette\@versim>}}
 \def\@versim#1#2{\vcenter{\offinterlineskip
 \ialign{$\m@th#1\hfil##\hfil$\crcr#2\crcr\sim\crcr } }}
 \catcode`\@=12

 \catcode`@=12
\begin{document}
 \thispagestyle{empty}
 \begin{flushright}
 UCRHEP-T589\\
 Apr 2018\
 \end{flushright}
 \vspace{0.6in}
 \begin{center}
 {\LARGE \bf Dark SU(2) Antecedents\\ of the U(1) Higgs Model\\}
 \vspace{1.2in}
 {\bf Ernest Ma\\}
 \vspace{0.2in}
{\sl Physics and Astronomy Department,\\ 
University of California, Riverside, California 92521, USA\\}
\vspace{0.1in}
{\sl Jockey Club Institute for Advanced Study,\\ 
Hong Kong University of Science and Technology, Hong Kong, China\\} 
\end{center}
 \vspace{1.2in}

\begin{abstract}\
The original spontaneously broken $U(1)$ gauge model with one complex Higgs 
scalar field has been known in recent years as a possible prototype 
dark-matter model.  Its antecedents in the context of $SU(2)$ are discussed.
Three specific examples are described, with one dubbed 
``quantum scotodynamics''.
\end{abstract}

 \newpage
 \baselineskip 24pt

\noindent \underline{\it Introduction}~:~
Consider the addition of the $U(1)_D$ Higgs model~\cite{h64} to the standard 
$SU(3)_C \times SU(2)_L \times U(1)_Y$ gauge model (SM) of quarks and leptons. 
The former may be used for dark matter~\cite{llm12,fr12,bkps13,dft16,dm17,m17} 
because it has the built-in $Z_2$ symmetry where the massive gauge boson 
$Z_D$ after spontaneous symmetry breaking is odd and the one physical real 
scalar boson $h_D$ is even.  However, $U(1)_D$ may mix kinetically~\cite{h86} 
with $U(1)_Y$, in which case the above $Z_2$ symmetry would be violated. 
To avoid this problem, it is suggested here that $U(1)_D$ be replaced with 
an $SU(2)$ antecedent, with an enriched dark-matter sector.  Three explicit 
examples will be discussed.  Note that this version of dark $SU(2)$ requires 
that it be broken to $U(1)$, in contrast to the case where a local or global 
$SU(2)$ dark symmetry remains~\cite{h09}.

\noindent \underline{\it $SU(2)_D$ with Scalar Doublet and Triplet}~:~
To break $SU(2)_D$ to $U(1)_D$, the simplest choice is a real scalar 
triplet
\begin{equation}
\chi = (\chi_1, \chi_2, \chi_3)
\end{equation}
with $\langle \chi_3 \rangle = v_3$.  In that case, the vector gauge bosons
\begin{equation}
W_D^\pm = {D_1 \pm iD_2 \over \sqrt{2}}
\end{equation}
acquire mass given by $m^2_{W_D} = 2g_D^2 v_3^2$. Note that the superscript 
$\pm$ refers to dark charge, the details of which will be discussed later.

To break $U(1)_D$ in the context of $SU(2)_D$ so that $D_3 = Z_D$ acquires 
mass, a complex scalar doublet
\begin{equation}
\Phi = \pmatrix{\phi_1 \cr \phi_2}
\end{equation}
is used.  Moreover, a global $U(1)_\Phi$ symmetry is imposed, i.e.
\begin{equation}
\Phi \to e^{i \theta} \Phi,
\end{equation}
which prevents the coupling of $\vec{\chi}$ to the triplet 
$\phi_i \epsilon_{ij} \vec{\sigma}_{jk} \phi_k$. 
The scalar potential consisting of $\chi$ and $\Phi$ is then given by
\begin{eqnarray}
V &=& m_2^2 \Phi^\dagger \Phi + {1 \over 2} m_3^2 (\vec{\chi} \cdot \vec{\chi})
 + \mu_0 \Phi^\dagger (\vec{\sigma} \cdot \vec{\chi}) \Phi \nonumber \\ 
&+& {1 \over 2} \lambda_2 (\Phi^\dagger \Phi)^2 + {1 \over 2} \lambda_3 
(\vec{\chi} \cdot \vec{\chi})^2 + \lambda_4 (\Phi^\dagger \Phi)(\vec{\chi} 
\cdot \vec{\chi}).
\end{eqnarray}
Note that the triplet combination of two identical real scalar triplets is 
zero.  The minimum of $V$ admits a solution
\begin{equation}
\langle \chi_{1,2} \rangle = 0, ~~~ \langle \chi_3 \rangle = v_3, ~~~ 
\langle \phi_1 \rangle = 0, ~~~ \langle \phi_2 \rangle = v_2/\sqrt{2},
\end{equation}
where $v_2$ is assumed real without any loss of generality, and
\begin{eqnarray}
0 &=& v_3[m_3^2 + 2\lambda_3 v_3^2 +  \lambda_4 v_2^2] - \mu_0 v_2^2/2, \\ 
0 &=& v_2[m_2^2 + \lambda_2 v_2^2/2 + \lambda_4 v_3^2 - \mu_0 v_3],
\end{eqnarray}
provided that
\begin{eqnarray}
&& m_2^2 + \lambda_4 v_3^2 + \mu_0 v_3 > 0, \\ 
&& m_2^2 + \lambda_4 v_3^2 - \mu_0 v_3 < 0.
\end{eqnarray}
As a result
\begin{equation}
m^2_{W_D} = 2g_D^2 v_3^2 + {1 \over 4} g_D^2 v_2^2, ~~~ 
m^2_{Z_D} = {1 \over 4} g_D^2 v_2^2, ~~~ m^2_{\phi_1} = 2 \mu_0 v_3,
\end{equation}
and the $2 \times 2$ mass-squared matrix spanning 
$h_D = \sqrt{2}Re(\phi_2)-v_2$ and $H_D = \chi_3 - v_3$ is given by
\begin{equation}
{\cal M}^2_{h_D,H_D} = \pmatrix{\lambda_2 v_2^2 & 
 v_2(2 \lambda_4 v_3 - \mu_0) \cr v_2(2 \lambda_4 v_3 - \mu_0)
 & 4 \lambda_3 v_3^2 + \mu_0 v_2^2/2v_3}.
\end{equation}
A global residual symmetry remains, under which 
\begin{equation}
W^+_D,\phi_1 \sim +1, ~~~ W^-_D,\phi_1^* \sim -1, ~~~ Z_D,h_D,H_D \sim 0.
\end{equation}
This comes from $I_{3D} + S_\Phi$, where $S_\Phi = 1/2$ for $\Phi$ and 
zero for all other fields.  It is possible because of the imposed global 
$U(1)_\Phi$ symmetry.  Whereas $\langle \phi_2 \rangle = v_2/\sqrt{2}$ 
breaks both 
$I_{3D}$ and $S_\Phi$, the linear combination $I_{3D}+S_\Phi$ is zero for 
$\phi_2$, so it remains as a residual dark symmetry. 

An important consequence of this structure is the emergence of a dark 
charge conjugation symmetry as in the original Higgs model~\cite{h64}, i.e.
\begin{equation}
W_D^+ \leftrightarrow W_D^-~(D_2 \leftrightarrow -D_2), ~~~ 
\phi_1 \leftrightarrow \phi_1^*, ~~~ Z_D (D_3) \leftrightarrow -Z_D (D_3).
\end{equation}
This comes from the gauge-invariant terms
\begin{equation}
-{1 \over 4} (\partial_\mu \vec{D}_\nu - \partial_\nu \vec{D}_\mu + 
g_D \vec{D}_\mu \times \vec{D}_\nu)^2 + |\partial_\mu \Phi - {i g_D \over 2} 
\vec{\sigma} \cdot \vec{D}_\mu \Phi|^2.
\end{equation}
It means that $Z_D$ is stable if its mass is less than twice that of 
$\phi_1$, in complete analogy to the $U(1)_D$ model of Ref.~\cite{m17}.
This makes it possible in principle to implement the inception of 
self-interacting dark matter, i.e. $\phi_1$ or $W_D$ of order 100 GeV 
with $Z_D$ as the light \underline{stable} mediator of order 10 to 100 MeV, 
to explain~\cite{fkty09} the observed core-cusp 
anomaly in dwarf galaxies~\cite{dgsfwggkw09}.  If $Z_D$ is unstable and 
decays to SM particles, as is the case for the light mediator proposed 
in most models, then very strong constraints exist~\cite{gibm09} from 
the cosmic microwave background (CMB) which 
basically rule out~\cite{bksw17} this scenario.  On the other hand, $h_D$ 
must also be light and decay quickly through its mixing with the SM Higgs 
boson $h$ before big bang nucleosynthesis (BBN).  In that case, the 
elastic scattering of $W_D$ or $\phi_1$ off nuclei through $h_D$ exchange 
is much too large to be acceptable with present data.  In Ref.~\cite{m17}, 
this is not a problem because the dark matter is a Dirac fermion which 
couples to $Z_D$ but not $h_D$.

As it is, this specific $SU(2)_D$ antecedent of the $U(1)$ Higgs model 
may still be a model of dark matter without addressing the core-cusp 
anomaly in dwarf galaxies.  Assuming that $W_D$ is heavy enough to decay 
into $\phi_1 h_D$ and $Z_D$ heavy enough to decay into $\phi_1 \phi_1^*$, 
then the complex scalar $\phi_1$ may be considered dark matter.  Assuming 
that $h_D$ is lighter than $\phi_1$, the annihilation cross section 
of $\phi_1 \phi_1^*$ at rest $\times$ relative velocity is given by
\begin{equation}
\sigma(\phi_1 \phi_1^* \to h_D h_D) ~ v_{rel} = {\lambda_2^2 
\sqrt{1-r_1} \over 
64 \pi m^2_{\phi_1}} \left[ 1 + {r_1 (2+r_1) \over (2-r_1)(4-r_1)} \right]^2,
\end{equation}
where $r_1 = m^2_{h_D}/m^2_{\phi_1}$.  Assuming as an example 
$m_{\phi_1} = 150$ GeV and $m_{h_D} = 100$ GeV, the above may be set 
equal to $4.4 \times 10^{-26}~cm^3/s$ for $\lambda_2 = 0.126$.

There is always the allowed quartic $\lambda_{2h}$ coupling between 
the $SU(2)_D$ Higgs doublet and the $SU(2)_L \times U(1)_Y$ Higgs 
doublet of the SM, so that $\phi_1$ interacts with quarks 
through the SM Higgs boson $h$ in direct-search experiments.  Using 
present data~\cite{xenon17}, it has been shown~\cite{kmppz18} that 
$\lambda_{2h} < 4.4 \times 10^{-4}$.  This is also the mixing bewteen 
$h_D$ and $h$.  Even with this limit on $\lambda_{2h}$, it can still 
be large enough so that $h_D$ decays promptly to $b \bar{b}$ in the 
early Universe.  This interaction~\cite{m17} also keeps 
$h_D$ in thermal equilibrium with the particles of the SM.

\noindent \underline{\it $SU(2)_D$ with Fermion Doublet}~:~
Consider the addition of a fermion doublet
\begin{equation}
\Psi = \pmatrix {\psi_1 \cr \psi_2}_L 
\end{equation}
to the $SU(2)_D$ model discussed in the previous section.  It has the 
allowed interactions
\begin{equation}
i \bar{\Psi} \gamma^\mu (\partial_\mu - {i g_D \over 2} \vec{\sigma} \cdot 
\vec{D}_\mu) \Psi + [f \tilde{\Psi} (\vec{\sigma} \cdot \vec{\chi}) \Psi 
+ H.c.],
\end{equation}
where $\tilde{\Psi} = (\psi_2,-\psi_1)_L$.  Since 
$\langle \chi_3 \rangle = v_3$, this shows that $\psi_{1,2}$ combine to form 
a Dirac fermion of mass $fv_3$.  To be specific, let $\psi_{1L}$ be the 
left-handed component of the Dirac fermion $\psi$, and $\psi_{2L}$ redefined 
as the conjugate of its right-handed component, i.e. 
$\psi_{2L} \sim \bar{\psi}_R$.  Now $\psi_1$ has dark charge 1/2 and 
$\psi_2$ has dark charge $-1/2$.  Together they form a Dirac fermion $\psi$ 
of charge 1/2, which interacts vectorially with $D_3 = Z_D$.  Note that 
$\bar{\psi} \gamma_\mu \psi$ is odd under dark charge conjugation as 
expected.  Note also that $\psi$ has no direct coupling to $h_D$ because 
of $SU(2)_D$ gauge invariance.  This allows the inception of self-interacting 
dark matter as described below.

Consider the elastic scattering of $\psi$ with $\bar{\psi}$ through the 
exchange of 
the light mediator $Z_D$.  Its cross section in the limit of zero 
momentum is
\begin{equation}
\sigma(\psi \bar{\psi} \to \psi \bar{\psi}) = {g_D^4 m^2_{\psi} \over 
64 \pi m^4_{Z_D}} = {m^2_{\psi} \over 4 \pi v_2^4}.
\end{equation}
For the benchmark value of $\sigma/m_{\phi_1} \sim 1~cm^2/g$ for 
self-interacting matter, this is satisfied for example with
\begin{equation}
m_{\psi} = 100~{\rm GeV}, ~~~ v_2 = 200~{\rm MeV}.
\end{equation}
This low-energy effective theory consisting of $\psi$, $Z_D$ and $h_D$ 
may be dubbed \underline{quantum} \underline{scotodynamics}, from the 
Greek 'scotos' meaning darkness.

Consider now the annihilation of $\psi \bar{\psi} \to Z_D Z_D$.  
Since $Z_D$ is much lighter than $\psi$, this cross section $\times$ 
relative velocity is given by
\begin{equation}
\sigma(\psi \bar{\psi} \to Z_D Z_D) ~ v_{rel} = {g_D^4  
 \over 256 \pi m^2_{\psi}}.
\end{equation}
For $m_{\psi} = 100$ GeV, and setting 
$\sigma v_{rel} = 4.4 \times 10^{-26}~cm^3/s$,
\begin{equation}
g_D = 0.42
\end{equation}
is obtained, which implies from Eq.~(11) that
\begin{equation}
m_{Z_D} = 42~{\rm MeV}.
\end{equation}
As shown in Ref.~\cite{m17}, the light mediator $Z_D$ is 
stable but annihilates quickly to $h_D$ which decays.  The cross section 
$\times$ relative velocity is given by
\begin{eqnarray}
\sigma(Z_D Z_D \to h_D h_D) ~ v_{rel} = 
{g_D^4 \sqrt{1-r} \over 64 \pi m^2_{Z_D}}  
\left[ {4[r^2+4(2-r)^2] \over 
(4-r)^2} - {24r(2+r) \over 9(2-r)(4-r)} + {8(2+r)^2 \over 9(2-r)^2} 
\right],
\end{eqnarray}
where $r = m^2_{h_D}/m^2_{Z_D}$. Assuming $m_{h_D} = 21$ MeV as an example 
so that $r=0.25$, the above is equal to $2 \times 10^{-18}~cm^3/s$, 
which is orders of magnitude greater than what is required for $Z_D$ to be 
a significant component of dark matter.  It may re-emerge at late times 
by $\phi_1 \phi_1^*$ annihilation through Sommerfeld enhancement, but its 
fraction as dark matter remains negligible.  Since $Z_D$ is stable, it would 
also not disturb~\cite{gibm09,bksw17} the cosmic microwave background (CMB).

As for $h_D$, it is allowed to mix with the SM Higgs boson $h$ in the 
$2 \times 2$ mass-squared matrix
\begin{equation}
{\cal M}^2_{h_D,h} = \pmatrix{\lambda_2 v_2^2 & \lambda_{2h} v_2 v_h \cr 
\lambda_{2h} v_2 v_h & m_h^2},
\end{equation}
where $v_h = 246$ GeV and $m_h = 125$ GeV.  For $m_{h_D} << m_h$, the 
$h_D-h$ mixing is $\theta_{2h} = \lambda_{2h} v_2 v_h/m_h^2$.  Assuming 
\begin{equation}
\lambda_{2h} = 0.01,
\end{equation}
then $\theta_{2h} = 3.8 \times 10^{-5}$ and the $h_D$ lifetime for 
$e^-e^+$ decay is given by
\begin{equation}
\Gamma^{-1}(h_D \to e^- e^+) = {8 \pi v_h^2 \over m_{h_D} m_e^2 \theta_{2h}^2} 
= 0.13 ~s,
\end{equation}
which is short enough not to affect big bang nucleosynthesis (BBN).
The decay of the SM Higgs boson to $h_D h_D$ is given by
\begin{equation}
\Gamma (h \to h_D h_D) = {\lambda_{2h}^2 v_h^2 \over 16 \pi m_h} = 
0.96~{\rm MeV},
\end{equation}
which is less than 25\% of the SM width of 4.12 MeV and allowed by present 
data.  Note that $\lambda_2 = 0.008$ in Eq.~(25) for $m_{h_D} = 21$ 
MeV.  Note also the important fact that $\psi$ does not couple directly 
to $h_D$, otherwise Eq.~(26) would be impossible, as discussed in the 
previous section.

In summary, a successful description of self-interacting fermion dark matter 
($\psi$ with $m_{\psi} = 100$ GeV) through a stable light vector gauge 
boson ($Z_D$ with $m_{Z_D} = 42$ MeV) in an $SU(2)_D$ gauge model has been 
rendered.  The Higgs scalar $h_D$ associated with $Z_D$ is also light 
(21 MeV), but it decays away quickly before the onset of BBN. Other heavier 
particles in the dark sector are $W_D^\pm$ (which decays to 
$\psi_1 \psi_1/ \psi_2 \psi_2$),  $\phi_1$ (which decays to $W_D^+ h_D$), 
and $H_D$ which mixes slightly with $h$ and $h_D$.

\noindent \underline{\it $SU(2)_D$ with Scalar Doublet and Quintet}~:~
In the previous two examples, an imposed symmetry of the $SU(2)_D$ 
scalar doublet $\Phi$, i.e. Eq.~(4), is necessary for obtaining a 
dark symmetry.  Hence the latter is not predestined~\cite{m18}, 
i.e. not the automatic consequence of gauge symmetry and particle 
content.  To have a predestined dark $Z_2$ symmetry, the simpler scalar 
triplet is now replaced with a scalar quintet.  This is analogous to 
having a fermion quintet~\cite{cfs06} in the SM for minimal dark 
matter.

Consider thereby the real scalar quintet
\begin{equation}
\zeta = (\zeta^{++},\zeta^+,\zeta^0,\zeta^-,\zeta^{--})
\end{equation}
with $\langle \zeta^0 \rangle = v_5$, then $W_D^\pm$ obtains a 
mass given by $m^2_{W_D} = 6g_D^2 v_5^2$ from absorbing $\zeta^\pm$. 
This leaves $\zeta^{\pm \pm}$ as physical scalar bosons with two units 
of dark charge, interacting with $Z_D$.  The scalar potential consisting 
of $\zeta$ and $\Phi$ is then given by
\begin{equation}
V = m_2^2 \Phi^\dagger \Phi + {1 \over 2} m_5^2 \zeta^\dagger \zeta 
+ {1 \over 2} \lambda_2 (\Phi^\dagger \Phi)^2 + \lambda_5 (\Phi^\dagger \Phi) 
(\zeta^\dagger \zeta)^2 + V_3 + V_4,
\end{equation}
where $V_3$ contains the one cubic invariant formed out of 3 scalar quintets 
and $V_4$ contains two quartic invariants.  Note that the triplet combination 
of two identical real scalar quintets is zero.  As a result, this scalar 
potential automatically has an extra $U(1)_\Phi$ symmetry, so that 
$I_{3D} + S_\Phi$ remains unbroken as $\phi_2$ acquires a vacuum 
expectation value $v_2/\sqrt{2}$ as explained previously.

Assuming that
\begin{equation}
m_\zeta < 2m_{\phi_1} < m_{Z_D} < m_{W_D},
\end{equation}
then $W^+_D$ decays to $\phi_1 h_D$, $Z_D$ decays to $\phi_1 \phi_1^*$, 
but both $\phi_1$ and $\zeta$ are stable.  Hence this is an explicit example 
of two-component dark matter under one dark $U(1)$ symmetry.  Let
\begin{equation}
m_\zeta = 200~{\rm GeV}, ~~~ m_{\phi_1} = 150~{\rm GeV}, ~~~ m_{h_D} = 
100~{\rm GeV},
\end{equation}
then using Eq.~(16) for $\sigma_1(\phi_1 \phi_1^* \to h_D h_D) v_{rel}$ 
and the analogous
\begin{eqnarray}
\sigma_2(\zeta \zeta^* \to h_D h_D, \phi_1 \phi_1^*) ~ v_{rel} &=& 
{\lambda_5^2 \sqrt{1-r_2} 
\over 64 \pi m^2_\zeta} \left[ 1 + {2(\lambda_5/\lambda_2)r_2 \over 2-r_2} 
-{3r_2 \over 4-r_2} \right]^2 \nonumber \\ 
&+& {\lambda_5^2 \sqrt{1-r_3} \over 32 \pi m^2_\zeta} \left[ 1 - {r_2 \over 
4-r_2} \right]^2,
\end{eqnarray}
where $r_2 = m^2_{h_D}/m^2_\zeta$ and $r_3 = m^2_{\phi_1}/m^2_\zeta$, 
the condition for the correct relic abundance is roughly given by
\begin{equation}
\langle \sigma_1 v_{rel} \rangle^{-1} + \langle \sigma_2 v_{rel} \rangle^{-1} 
= (4.4 \times 10^{-26}~cm^3/s)^{-1}. 
\end{equation}
It has for example the reasonable solution $\lambda_5 = \lambda_2 = 0.173$, 
in which case $\phi_1$ is 53\% and $\zeta$ 47\% of dark matter. 
Again the mixing of $\zeta$ with the SM Higgs boson $h$ must be small 
as it is for $\phi_1$ to satisfy direct-search limits as discussed 
previously.

In this scenario, the addition of the fermion doublet of Eq.~(17) could 
also provide a low-energy effective theory of quantum scotodynamics 
with light $Z_D$ and $h_D$.  
In that case, $\zeta^{\pm \pm}$ would decay into $W_D^\pm W_D^\pm$, 
$\phi_1$ would decay into $W_D^+ h_D$, and $W_D^\pm$ would decay into 
$\psi_1 \psi_1/\psi_2 \psi_2$.

\noindent \underline{\it Concluding Remarks}~:~
Exploring the possible $SU(2)$ antecedents of the famous $U(1)$ Higgs 
model for a nontrivial application to dark matter, three interesting 
examples have been identified and discussed.  The minimal version with 
one real scalar triplet $\chi$ and one complex scalar doublet $\Phi$ 
admits $\phi_1$ as dark matter, but a global $U(1)$ symmetry has to be 
imposed.  With the addition of a fermion doublet $\psi$, the inception 
of self-interacting dark matter may be implemented successfully, avoiding 
all potential astrophysical and laboratory constraints.  A third example 
replaces $\chi$ with the real scalar quintet $\zeta$, in which case 
the dark $U(1)$ symmetry becomes predestined, i.e. automatic from the gauge 
symmetry and particle content.

\noindent \underline{\it Acknowledgement}~:~
This work was supported in part by the U.~S.~Department of Energy Grant 
No. DE-SC0008541.

\bibliographystyle{unsrt}

\begin{thebibliography}{99}
\bibitem{h64} P. W. Higgs, Phys. Rev. Lett. {\bf 13}, 508 (1964).
\bibitem{llm12} O. Lebedev, H. M. Lee, and Y. Mambrini, Phys. Lett. 
{\bf B707}, 570 (2012).
\bibitem{fr12} Y. Farzan and A. Rezaei Akbarrieh, JCAP {\bf 1210}, 026 (2012).
\bibitem{bkps13} S. Baek, P. Ko, W.-I. Park, and E. Senaha, JHEP {\bf 1305}, 
036 (2013).
\bibitem{dft16} A. DiFranzo, P. J. Fox, and T. M. P. Tait, JHEP {\bf 1604}, 
135 (2016).
\bibitem{dm17} A. DiFranzo and G. Mohlabeng, JHEP {\bf 1701}, 080 (2017). 
\bibitem{m17} E. Ma, Phys. Lett. {\bf B772}, 442 (2017).
\bibitem{h86} B. Holdom, Phys. Lett. {\bf B166}, 196 (1986).
\bibitem{h09} T. Hambye, JHEP {\bf 0901}, 028 (2009).
\bibitem{fkty09} See for example J. L. Feng, M. Kaplinghat, H. Tu, and H.-B. 
Yu, JCAP {\bf 0907}, 004 (2009).
\bibitem{dgsfwggkw09} See for example F. Donato, G. Gentile, P. Salucci, C. 
Frigerio Martins, M. I. Wilkinson, G. Gilmore, E. K. Grebel, A. Koch, and 
R. Wyse, MNRAS {\bf 397}, 1169 (2009).
\bibitem{gibm09} S. Galli, F. Iocco, G. Bertone, and A. Melchiorri, Phys. Rev. 
{\bf D80}, 023505 (2009).
\bibitem{bksw17} T. Bringmann, F. Kahlhoefer, K. Schmidt-Hoberg, and P. Walia, 
Phys. Rev. Lett. {\bf 118}, 141802 (2017).
\bibitem{xenon17} E. Aprile, {\it et al.}, XENON Collaboration, Phys. Rev. 
Lett. {\bf 119}, 181301 (2017).
\bibitem{kmppz18} C. Kownacki, E. Ma, N. Pollard, O. Popov, and M. Zakeri, 
Eur. Phys. J. {\bf C78}, 148 (2018).
\bibitem{m18} E. Ma, arXiv:1803.03891 [hep-ph].
\bibitem{cfs06} M. Cirelli, N. Fornengo, and A. Strumia, Nucl. Phys. 
{\bf B753}, 178 (2006).
\end{thebibliography}

\end{document}